\documentclass{emulateapj}
\usepackage{epstopdf}
\usepackage{longtable}

\newcommand {\aplt} {\ {\raise-.5ex\hbox{$\buildrel<\over\sim$}}\ } 
\newcommand {\apgt} {\ {\raise-.5ex\hbox{$\buildrel>\over\sim$}}\ }

\slugcomment{Accepted for Publication in ApJ}
\shorttitle{iPTF 16bco}
\shortauthors{Gezari et al.}

\begin{document}

\title{\lowercase{i}PTF Discovery of the Rapid "Turn On" of a Luminous Quasar}

\author{S.~Gezari\altaffilmark{1,2}, T.~Hung\altaffilmark{1}, S.~B.~Cenko\altaffilmark{3,2}, N. Blagorodnova\altaffilmark{4}, Lin Yan\altaffilmark{5,6}, S.~R.~Kulkarni\altaffilmark{4}, K.~Mooley\altaffilmark{7}, A.~K.~H.~Kong\altaffilmark{8}, T.~M.~Cantwell\altaffilmark{9}, P.~C.~Yu\altaffilmark{10}, Y.~Cao\altaffilmark{11}, C. Fremling\altaffilmark{12}, J.~D.~Neill\altaffilmark{4}, C.-C.~Ngeow\altaffilmark{10}, P.~E.~Nugent\altaffilmark{13,14}, and P.~Wozniak\altaffilmark{15}}
\altaffiltext{1}{Department of Astronomy, University of Maryland, Stadium Drive, College Park, MD 20742-2421, USA \email{suvi@astro.umd.edu}}
\altaffiltext{2}{Joint Space-Science Institute, University of Maryland, College Park, MD 20742, USA}
\altaffiltext{3}{NASA Goddard Space Flight Center, Mail Code 661, Greenbelt, MD 20771, USA}
\altaffiltext{4}{Department of Astronomy, California Institute of Technology, Pasadena, CA 91125, USA}
\altaffiltext{5}{Caltech Optical Observatories, Cahill Center for Astronomy and Astrophysics, California Institute of Technology, Pasadena, CA 91125, USA}
\altaffiltext{6}{Infrared Processing and Analysis Center, California Institute of Technology, Pasadena, CA 91125, USA}
\altaffiltext{7}{Astrophysics, Department of Physics, University of Oxford, Keble Road, Oxford OX1 3RH, UK}
\altaffiltext{8}{Institute of Astronomy, National Tsing Hua University, Hsinchu 30013, Taiwan}
\altaffiltext{9}{Jodrell Bank Centre for Astrophysics, Alan Turing Building, Oxford Road, Manchester M13 9PL, UK}
\altaffiltext{10}{Graduate Institute of Astronomy, National Central University, Taoyuan City 32001, Taiwan}
\altaffiltext{11}{eScience Institute and Astronomy Department, University of Washington, Seattle, WA 98195, USA}
\altaffiltext{12}{Department of Astronomy, The Oskar Klein Center, Stockholm University, AlbaNova, 10691, Stockholm, Sweden}
\altaffiltext{13}{Department of Astronomy, University of California, Berkeley, CA 94720-3411, USA}
\altaffiltext{14}{Lawrence Berkeley National Laboratory, 1 Cyclotron Road, MS 50B-4206, Berkeley, CA 94720, USA}
\altaffiltext{15}{Los Alamos National Laboratory, MS D436, Los Alamos, NM, 87545, USA}

\begin{abstract}
We present a radio-quiet quasar at $z=0.237$ discovered ``turning on'' by the intermediate Palomar Transient Factory (iPTF).  The transient, iPTF 16bco, was detected by iPTF in the nucleus of a galaxy with an archival SDSS spectrum with weak narrow-line emission characteristic of a low-ionization emission line region (LINER).  Our follow-up spectra show the dramatic appearance of broad Balmer lines and a power-law continuum characteristic of a luminous ($L_{\rm bol} \approx 10^{45}$ ergs s$^{-1}$) type 1 quasar 12 years later.  Our photometric monitoring with PTF from 2009-2012, and serendipitous X-ray observations from the XMM-Newton Slew Survey in 2011 and 2015, constrain the change of state to have occurred less than 500 days before the iPTF detection.   An enhanced broad H$\alpha$ to [O~III] $\lambda$5007 line ratio in the type 1 state relative to other changing-look quasars also is suggestive of the most rapid change of state yet observed in a quasar.  We argue that the $> 10$ increase in Eddington ratio inferred from the brightening in UV and X-ray continuum flux is more likely due to an intrinsic change in the accretion rate of a pre-existing accretion disk, than an external mechanism such as variable obscuration, microlensing, or the tidal disruption of a star.  However, further monitoring will be helpful in better constraining the mechanism driving this change of state.  The rapid ``turn on'' of the quasar is much shorter than the viscous infall timescale of an accretion disk, and requires a disk instability that can develop around a $\sim 10^{8} M_\odot$ black hole on timescales less than a year.  \end{abstract}

\keywords{galaxies:active -- accretion, accretion disks -- black hole physics -- surveys}

\section{Introduction}

Variability is a ubiquitous property of quasars on timescales of hours to years, and likely attributed to processes in the accretion disk fueling the central supermassive black hole (SMBH) \citep{Pereyra2006, Kelly2009}.   More dramatic changes in accretion activity are also expected to occur on much longer timescales.  Hydrodynamic simulations of quasar fueling and feedback reveal a ``duty cycle'' of accretion activity on timescales of a Myr \citep{Novak2011}, and may explain the lack of clear observational evidence for causal connections between AGN activity and star-formation in galaxy studies \citep{Hickox2014}.  Observations of ionization nebulae on the outskirts of galaxies have indicated the shut off of quasar engines on timescales of tens of thousands of years \citep[e.g.,][]{Schawinski2010}, and our own Galactic center shows signs of X-ray reflection from enhanced nuclear activity on a timescale of only hundreds of years ago \citep{Ponti2010}.  However, as we explore the behavior of active galactic nuclei (AGN) more systematically in the time domain with optical imaging and spectroscopic surveys, we are finding evidence for significant accretion state changes on even shorter timescales.

In the AGN unification model, type 1 (spectra with narrow and broad lines) and type 2 (spectra with narrow lines only) classifications are explained as a viewing angle effect due to nuclear obscuration of the broad-line region and accretion disk continuum \citep{Antonucci1993}.  A challenge to this paradigm is the rare class of ``changing-look'' AGN, who change their spectral class with the appearance and/or disappearance of broad Balmer lines, accompanied by large-amplitude changes in the AGN continuum.  Most of the changing-look AGN cases reported to date have been spectroscopically known Seyferts who have shown a dramatic appearance or disappearance of their broad Balmer lines in follow-up spectra resulting in a ``change of state'' between a type 1 and a type 1.8-2 spectrum.  A type 1.8 or 1.9 Seyfert classification depends on the presence of weak broad H$\beta$ or broad H$\alpha$, respectively \citep{Osterbrock1981}.   

Notable examples of changing-look AGN include the appearance of broad, double-peaked H$\alpha$ and H$\beta$ lines in LINER galaxy NGC 1097 \citep{Storchi1993}, and the complete disappearance of the broad H$\beta$ line in Seyfert 1 galaxy Mrk 590 \citep{Denney2014}. 
More recently the ASAS-SN optical time domain survey discovered an outburst from NGC 2617 at $z=0.0142$, that was accompanied by a transition from a Seyfert 1.8 to a Seyfert 1 optical spectrum on a timescale of $\sim 10$ yr \citep{Shappee2014}.  And recent follow-up observations of the changing-look AGN Mrk 1018, which had changed from a type 1.9 to type 1 Seyfert in 1984, revealed a change back to a type 1.9 Seyfert 30 years later \citep{McElroy2016}.  The fact that the variable UV/optical continuum in Mrk 1018 followed the $L \sim T^4$ relation expected for thermal emission from a disk, and that there was no evidence for neutral hydrogen absorption in its X-ray spectrum, favored intrinsic changes in the accretion flow instead of an obscuration event \citep{Husemann2016}, which has been inferred to be the cause for rapid drops in the UV continuum \citep{Guo2016} and X-ray flux \citep{Risaliti2009, Marchese2012} in some AGN.  

The first case of a changing-look AGN with the luminosity of a quasar (which we define as $L_{\rm bol} > 10^{44}$ ergs s$^{-1}$) was SDSS J0159+0033 at $z=0.312$ \citep{LaMassa2015}, which was discovered to change from a type 1 to type 1.9 spectrum between its SDSS DR1 spectrum in 2001 and its SDSS-III BOSS spectrum in 2010.  Since then, there was a systematic search by \citet{Ruan2016} of SDSS quasars with multiple epochs of spectra for which the spectroscopic pipeline classification changed between a "QSO" to a "GALAXY", or vice versa, that recovered SDSS J0159+0033, and revealed two more cases of changing-look quasars (at $z=0.198$ and $z=0.243$) with a dimming in their continuum and disappearance of the broad H$\beta$ line on a timescale of $5-7$ years in the rest-frame.   An archival search of SDSS quasars with multiple epochs of spectra {\it and} large amplitudes of variability ($\Delta g > 1$ mag) by \citet{MacLeod2016} also yielded SDSS J0159+0033, and 9 more changing-look quasars at $z=0.2-0.6$, 5 of which show the {\it appearance} of broad H$\beta$ between the SDSS and BOSS spectral epochs.  New epochs of spectra from the Time-Domain Spectroscopic Survey (TDSS; \citet{Morganson2015}) have also revealed one new case of a transition from a type 1 to type 1.9 quasar at z=0.246 \citep{Runnoe2016}.  

Here we report the rapid ($< 1$ yr) emergence of a type 1 radio-quiet quasar from a galaxy at $z=0.237$ with weak narrow-line emission in its pre-event spectrum characteristic of a LINER nucleus that does not require an AGN to power the line emission.  Throughout this paper, we refer to quasars as radio quiet or radio loud AGN above a bolometric luminosity of $10^{44}$ ergs s$^{-1}$.  We also adopt a cosmology where $H_0 = 70$ km s$^{-1}$ and $\Omega_M = 0.3$ and $\Omega_\Lambda = 0.7$, yielding a luminosity distance for iPTF 16bco of $d_L = 1186$ Mpc.

\section{Observations}

\subsection{SDSS Archival Imaging and Spectroscopy}\label{sdss}

The source SDSS J155440.25+362952.0 was imaged by the SDSS survey on UT 2003 April 29 (all days hereafter are in the UT system), and morphologically classified as a galaxy with $r = 18.18 \pm 0.01$ mag.  It was targeted in the SDSS spectroscopic legacy survey as a $ugri$-selected quasar, selected for lying more than 4$\sigma$ from the stellar locus \citep{Richards2002} at high Galactic latitude (QSO\_CAP).  
It turns out that while the colors measured in the SDSS survey are outside the stellar locus, with $u-g = +0.44 \pm 0.04$ mag, $g-r = +0.92 \pm 0.01$ mag, and $r-i = +0.48 \pm 0.01$ mag, given the known redshift of the galaxy, they are inconsistent with the quasar color-redshift relation measured for the SDSS spectroscopic sample \citep{Schneider2007}.  

The SDSS legacy spectrum, obtained on 2004 June 16, was determined by the spectroscopic pipeline to have a $z=0.2368$, and a spectroscopic classification of a "GALAXY", with a stellar velocity dispersion of $\sigma_\star = 176 \pm 14$ km/s.  The galaxy classification was due the presence of strong galaxy absorption features (Ca H\&K, G band, Mg I, Na D), with only weak [O III] and [N~II] emission lines detected.  According to the $M_{\rm BH}-\sigma_\star$ scaling relation, this velocity dispersion corresponds to a central black hole mass of $(1^{+2}_{-0.7}) \times 10^8  M_\odot$ \citep{McConnell2013}.  

We find no evidence for significant flux variations between the SDSS survey image in 2003 and legacy spectrum in 2004.  The synthetic $r$-band magnitude of the 2004 SDSS spectrum (measured by projecting the best-fit spectral template onto the $r$-band filter) is {\tt spectroSynFlux\_r} $ = 18.82 \pm 0.02$ mag.  The corresponding fiber magnitude for the SDSS imaging in 2003 (measured with an aperture equal to the 3$\arcsec$ diameter spectroscopic fiber) is $r=18.95$ mag.  Since the SDSS spectrum in 2004 is dominated by host galaxy starlight in the wavelength range of the $r$-band ($\lambda_{\rm eff} = 6231$ \AA), we conclude that the SDSS image in 2003 is also dominated by host galaxy starlight.

\subsection{\textsl{GALEX} Archival Imaging} \label{sec:hst}
The source was observed by the \textsl{GALEX} All-Sky Imaging Survey on 2004 May 15 with an 6 $\arcsec$ (4 pixel) radius aperture magnitude corrected for the total energy enclosed \citep{Morrissey2007} of $NUV = 21.61 \pm 0.35$ mag, and a 5$\sigma$ point-source upper limit of $FUV > 20.6$ mag for $t_{\rm exp} = 112$ s, and a background of $2.6 \times 10^{-3}$ cts s$^{-1}$ pix$^{-1}$.  Note that the color measured by \textsl{GALEX} and SDSS of $NUV - r = +3.4 \pm 0.35$ mag is entirely consistent with normal galaxies with a similar luminosity on the blue sequence \citep{Wyder2007}.

\subsection{iPTF Detection} \label{sec:hst}
iPTF 16bco was discovered as a transient detection by the Palomar 48-in telescope (P48) on 2016 Jun 1 during an (intermediate) Palomar Transient Factory (iPTF) $g+r$ band experiment by the real-time difference imaging pipeline run at LBNL \citep{Cao2016}, with $g=19.4$ mag and $r=19.6$ mag.  The source was flagged as a ``nuclear'' transient given the measured offset of $0.44$ arcsec from its host galaxy; within our centroiding accuracy of $0.8$ arcsec.  The data were re-reduced with the PTFIDE pipeline run at IPAC \citep{Masci2016}.

\subsection{Follow-up Spectroscopy} 
On the next day after the discovery (2016 Jun 2), the transient iPTF 16bco was followed-up with the robotic low-resolution (R = $\lambda/\Delta \lambda \sim 100$) Integral Field Unit (IFU) spectrograph, part of the Spectral Energy Distribution Machine (SEDM) instrument on the Palomar 60-in telescope (P60).  The spectrum was strikingly different than its archival SDSS spectrum, and three more epochs of spectroscopy were obtained with DEIMOS \citep{Faber2003} on the Keck-II telescope on 2016 June 4, and the DeVeny spectrograph on the Discovery Channel Telescope (DCT) on 2016 June 13 and 2016 July 9.   

\begin{figure*}
\plotone{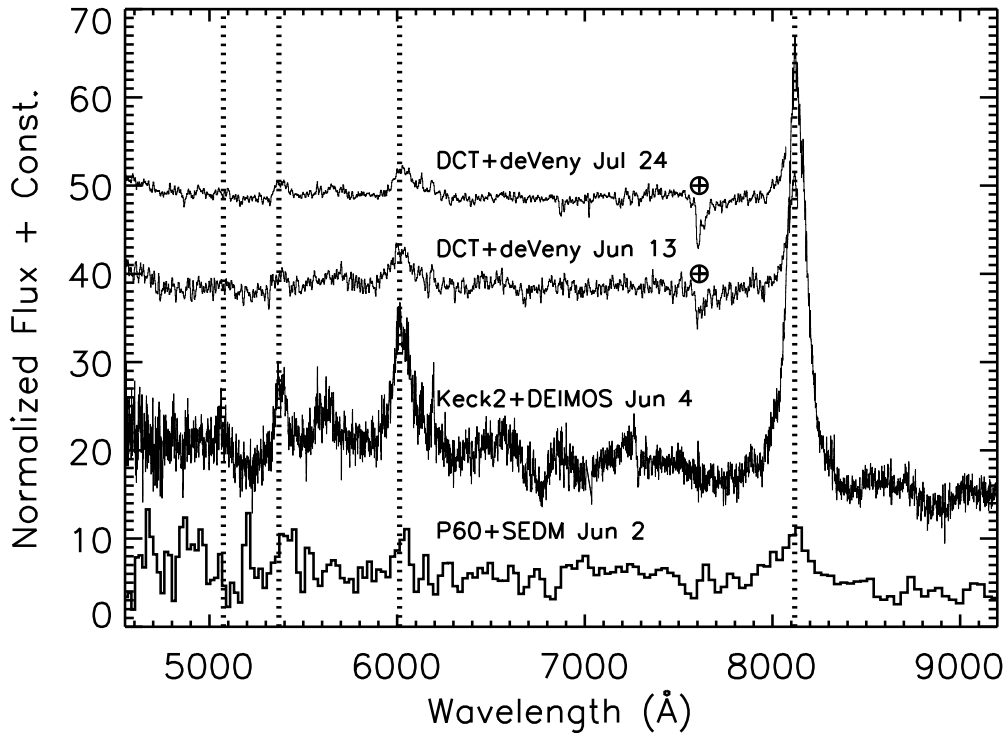}
\caption{Series of follow-up spectra of iPTF 16bco.  Spectra are normalized, and have been offset vertically for clarity.  Dotted lines show the wavelengths of broad Balmer lines at $z=0.2368$.  The uncorrected telluric A-band absorption feature at $\sim$ 7600 \AA~ in the DCT spectra are also marked.
 \label{fig:specall} }
\end{figure*}

The SEDM IFU obtained 2700 s exposures on 2016 June 2 and July 24, and data reduction was performed by the SEDM pipeline\footnotetext{http://www.astro.caltech.edu/sedm}.  Sky subtraction was performed using ``A-B'' extraction:  two exposures of the same length are taken offset by a few arcsec, so that the object of interest lies on an empty sky region on the next exposure, and then subtracted from each other to remove the sky lines.  The spectrum is extracted in each exposure separately, and then the fluxes are summed to get the final spectrum.

The Keck spectrum was obtained with a 0$\farcs$8 slit and the exposure time 
was 240 seconds. Data were reduced with usual procedures from the DEEP2 pipeline in 
IDL \citep{Cooper2012, Newman2013} and in PyRAF. The flux calibration and 
telluric correction were done with the flux standard star BD262606. The spectrum covers 
wavelengths ranging from 4550 \AA~ to 9550 \AA~with a spectral resolution of $\sim$4 \AA.  The DCT spectra were
obtained with a 1$\farcs$5 slit and an exposure time of 600 s and 1200 s, respectively, with a spectral resolution of $\sim 9$ \AA.  Data were reduced in standard IRAF routines, and flux calibration was performed with the flux standard star BD+40d4032.

Figure \ref{fig:specall} shows the series of follow-up spectra, which demonstrate strong, broad, Balmer emission lines characteristic of a type 1 quasar at $z=0.2368$.
 
\subsection{Palomar 60 inch Imaging} \label{sec:hst}
\begin{figure*}
\plotone{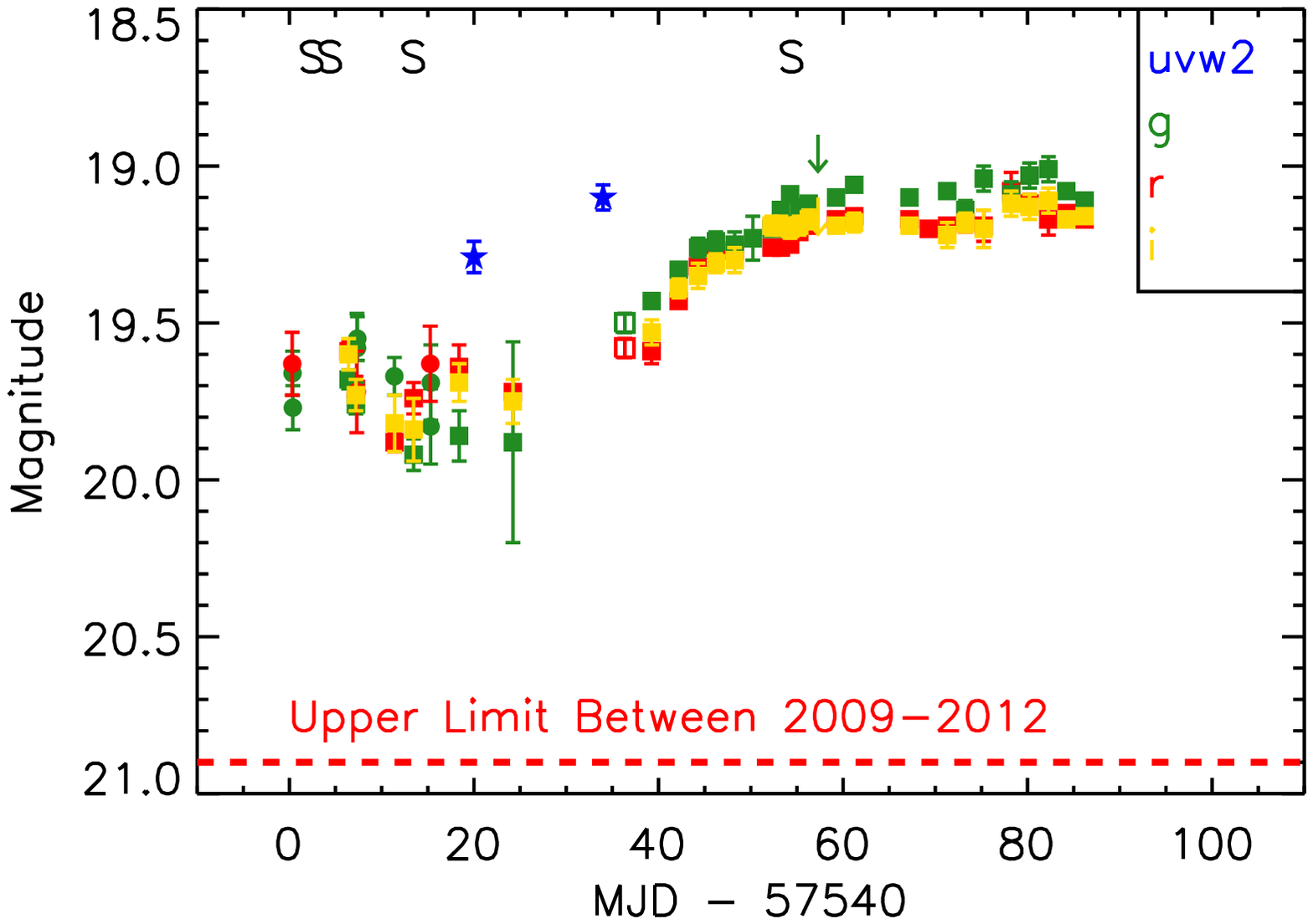}
\caption{Light curve of iPTF 16bco.  Optical $g$, $r$, and $i$ band difference-imaging photometry are from the Palomar 48-in (solid-circles) and 60-in telescopes (solid-squares: SEDM, open squares: GBMCam), while ultraviolet aperture photometry in the $uvw2$ band is from the {\it Swift} telescope (stars), with a negligible contribution from host galaxy light.  Dashed red line shows the mean $r$-band upper limit measured during PTF observations between $2009-2012$.   The epochs for which spectra were taken are marked with an S.
 \label{fig:lc} }
\end{figure*}

We also monitored the source in 3 filters ($g, r, i$) with the SEDM on the P60 telescope.  The data were host-subtracted using {\tt FPipe} \citep{Fremling2016}.  One epoch of the P60 data points on MJD 57576 are from the GRBCam.   We do not plot the $i$-band GRBCam data from this night due to the large difference in the shape of its filter transmission curve in this band.  The light curve of the transient iPTF 16bco is presented in Figure \ref{fig:lc}, and the photometry is given in Table \ref{tab1}.  We adjust the $g$-band P48 photometry by +0.25 mag in order to match the P60 photometry.  This offset is attributed to the difference in filter curves in the $g$-band, and the strong blue continuum.  Note that since its discovery by iPTF on 2016 Jun 1 in the $g$ and $r$ bands, iPTF 16bco has retained a blue color, $g-r \approx -0.1$ mag, and demonstrated a rise in brightness of $\sim$ 0.5 mag over a timescale of $\sim 1$ month.

\subsection{PTF Historical Light Curve} \label{sec:hst}
The source was observed in the PTF survey in $2009-2012$, and no variability was detected in this time frame, with a median transient point source upper limit over 151 epochs of $r < 20.9$ mag, with the last non detection on 2012 May 28.  When adding in the host galaxy flux measured by SDSS, this corresponds to a total magnitude of $r < 18.10$ mag, or a $\Delta r < 0.08$ mag during this time period.  The P48 observations constrain the onset of the nuclear transient to be after the last non-detection on 2012 May 28, 4 years before the iPTF discovery.  

\subsection{Swift Observations} \label{sec:hst}
The source was observed with our Swift Key Project program for UV follow-up of iPTF nuclear transients (PI: Gezari) on 2016 June 21 and July 5.  We extracted the source from a 5$\farcs$0 region using a background region of 20$\farcs$0 radius using the task \texttt{uvotsource} in HEASoft which includes a correction for the enclosed energy in the aperture.  
The source was observed in the $uvw2$ filter with $\sim 1$ ks exposures, and detected with $19.30 \pm 0.05$ mag and $19.11 \pm 0.05$ mag in the AB system, respectively.  The corresponding UV-optical color of iPTF 16bco (with a negligible contribution of UV flux from the host) is $NUV-r \sim `-0.5$ mag, notably bluer than the $NUV-r$ colors of low-redshift quasars measured by \textsl{GALEX} and SDSS ($NUV-r = 0.0-0.5$ mag) \citep{Bianchi2005, Agueros2005}.  

Simultaneous {\it Swift} XRT observations were processed with the UK
Swift Data Science Centre\footnote[1]{http://www.swift.ac.uk/user\_objects/} pipeline that takes into account dead columns and
vignetting to extract counts from the source in the energy range of
0.3-10 keV. The X-ray count rate on June 21 and July 5 is
$0.025\pm0.0057$ and $0.013\pm0.0059$ counts s$^{-1}$, respectively.
We further obtained a 1.7 ks {\it Swift} XRT exposure on 2016 October
21 and the source was detected with $0.027\pm0.004$ counts s$^{-1}$.
This confirms the lack of significant X-ray variability between the
{\it Swift} observations. Furthermore, the combined spectrum of all the data can be
modeled with an absorbed power-law with a spectral index of
$\Gamma=2.1\pm0.5$  and $N_H$ fixed at the Galactic value of $1.63\times10^{20}$ cm$^{-2}$ \citep{Dickey1990}. The
average unabsorbed 0.2-10 keV flux of the source is $9\times10^{-13}$
ergs s$^{-1}$ cm$^{-2}$, corresponding to a luminosity of
$1.5\times10^{44}$ ergs s$^{-1}$ by assuming an absorbed power-law
with $N_{\rm H}=1.63\times10^{20}$ cm$^{-2}$ and $\Gamma=2.1$ at $z=0.2368$.

\subsection{Archival X-ray Observations} \label{sec:xray}

The ROSAT upper limit in the 0.1-2.4 keV band from All-Sky Survey in 1990-1991 \citep{Voges1999} is 0.1 cts s$^{-1}$ which corresponds to an unabsorbed flux of $\sim 7 \times 10^{-13}$ ergs s$^{-1}$ cm$^{-2}$.  There are even more constraining 2$\sigma$ upper limits from the XMM Slew Survey\footnote[2]{http://xmm.esac.esa.int/UpperLimitsServer/} of $< 0.601$ and $< 0.817$ cts s$^{-1}$ in the $0.2-12.0$ keV band on 2011 Feb 27 and 2015 Feb 08 corresponding to $< 2.7 \times 10^{-13}$ and $< 3.2 \times 10^{-13}$ ergs s$^{-1}$ cm$^{-2}$, respectively.  The latest XMM Slew Survey upper limit  implies a factor of $> 3$ increase in flux in the {\it Swift} XRT detection on a timescale of $< 1.1$ yr in the quasar rest-frame.

\subsection{Radio Observations}

We observed iPTF 16bco with the AMI-LA at 15.5 GHz on 2016 Oct 16.63.  The source is not detected, with the 3$\sigma$ upper limit of 68 $\mu$Jy. We can also convert this to a 1.4 GHz upper limit of 370 $\mu$Jy using a spectral index of $-0.7$.  This is consistent with the non-detection in the VLA FIRST survey \citep{Becker1995} from 1999 which gives an independent 1.4 GHz upper limit of 500 $\mu$Jy. 

\section{Analysis} \label{sec:analysis}

\subsection{Host Galaxy Classification}
\begin{figure*}
\plottwo{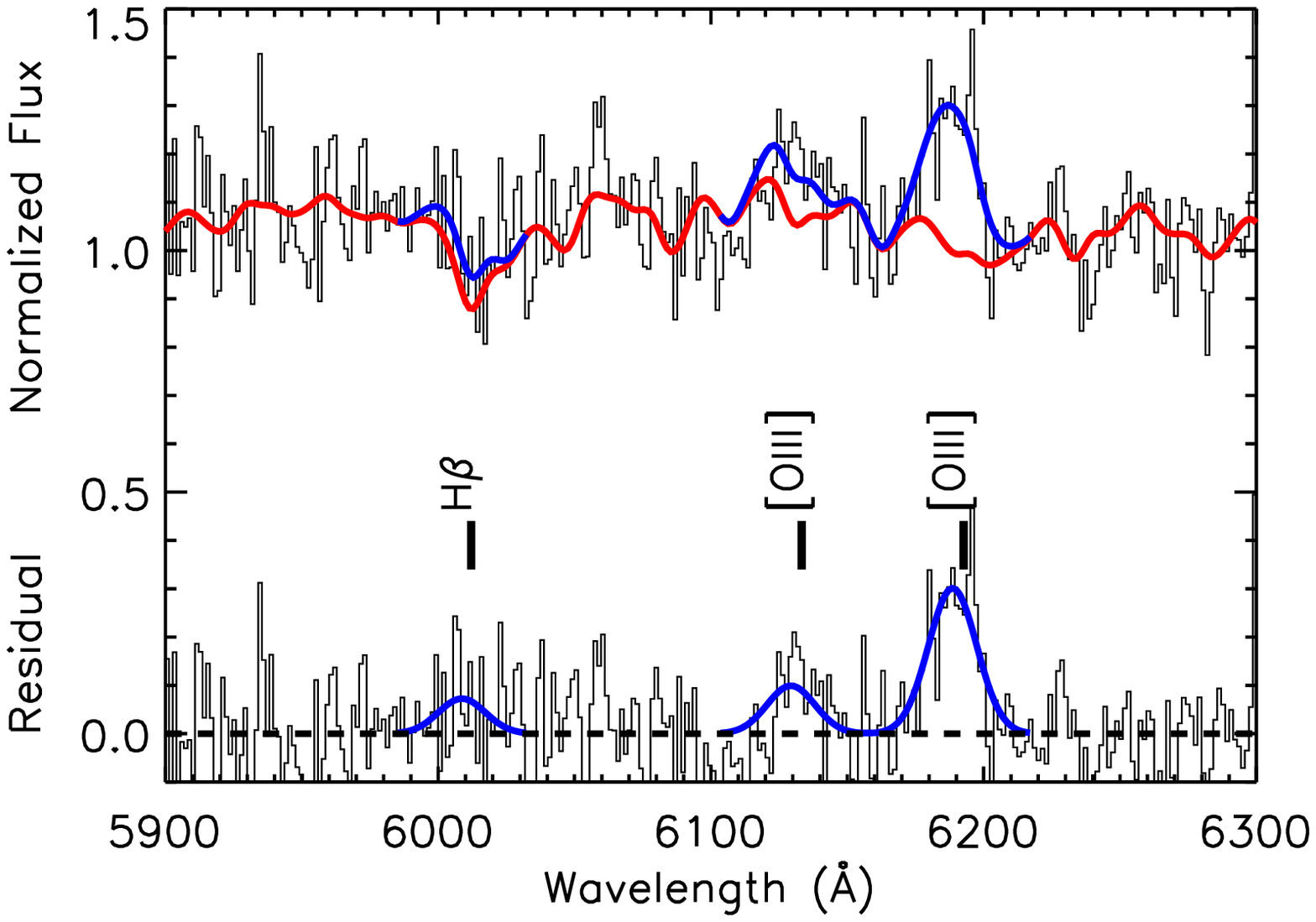}{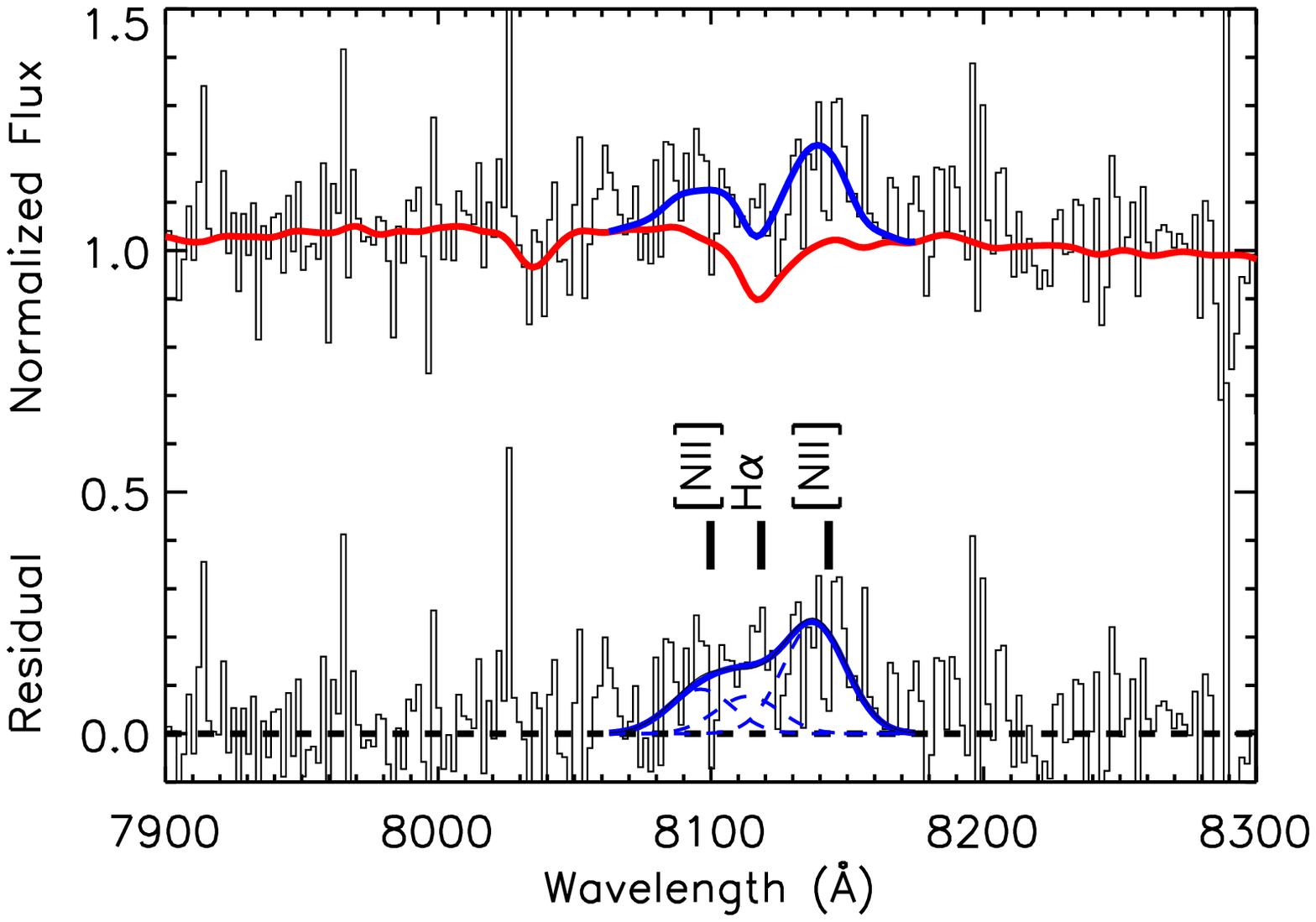}
\caption{H$\beta$ (left) and H$\alpha$ (right) regions of iPTF 16bco during its pre-event spectrum from SDSS.  The red line shows our galaxy template fit, and blue line shows the emission line component fit.  The residual from the galaxy template fit is also shown, along with the emission line component fit.  The individual Gaussian components of the [NII]+H$\alpha$ complex are plotted with dashed lines.  The emission lines are all fitted with a narrow Gaussian with a $\sigma = 420$ km s$^{-1}$, with no evidence for a broad H$\alpha$ or H$\beta$ line.   \label{fig:ha} }
\end{figure*}

\begin{figure*}
\plotone{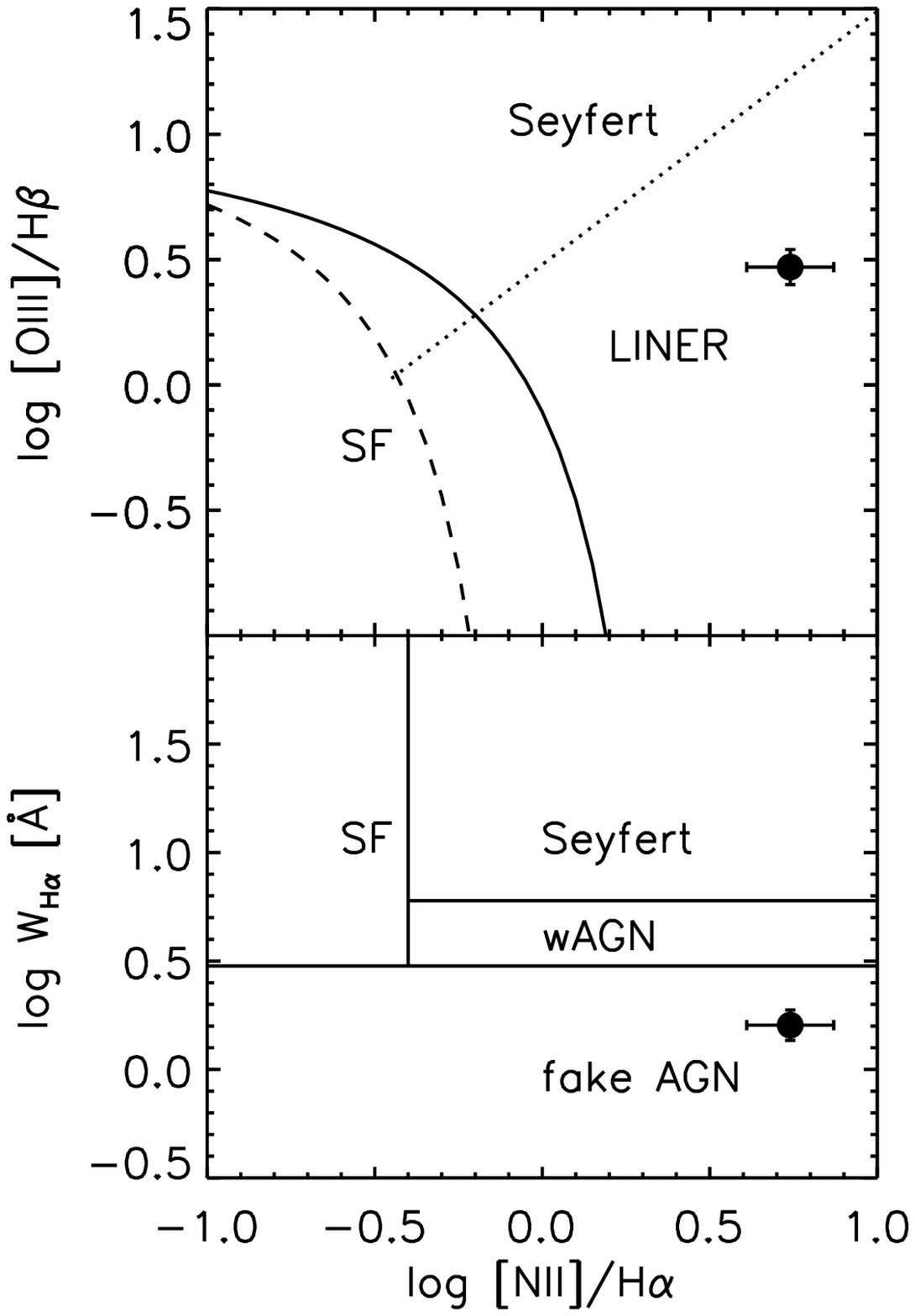}
\caption{Diagnostic narrow-line ratio diagrams for iPTF 16bco during its pre-event spectrum from SDSS.  {\it Left:} The BPT diagram diagram with the lines demarcating star-forming galaxies from AGN.  The solid line is the theoretical curve from \citet{Kewley2001}, and dashed line is the empirical curve based on the SDSS spectroscopic sample from \citet{Kauffmann2003}.  The line demarcating Seyferts from LINER galaxies from \citet{Cid2010} is plotted with a dotted line.  {\it Right:} The diagnostic WHAN diagram defined by \cite{Cid2010} with the regions demarcating star-forming galaxies, Seyferts, weak AGN (wAGN) and ``fake AGN'' powered by stars from \cite{Cid2011}.  \label{fig:bpt} }
\end{figure*}

The archival SDSS spectrum from 2004 was fitted by the automated spectroscopic pipeline \citep{Bolton2012} with a combination of stellar, galaxy, and quasar templates plus emission lines.  After visual inspection of the pipeline fit, we found a poor fit to the the H$\alpha$+[N~II] and [O~III] emission line complexes, and refitted the spectrum with host galaxy template and emission-line gas components using {\tt ppxf} \citep{Cappellari2004, Cappellari2017} which uses the MILES stellar template library \citep{Vazdekis2010}.  The emission line fits are shown in Figure \ref{fig:ha}, and are fitted with a narrow Gaussian with a $\sigma = 420$ km s$^{-1}$, with no evidence for a broad H$\alpha$ or H$\beta$ components.   The narrow-line ratios of $\log$([O~III]$\lambda 5007$/H$\beta)  = 0.74 \pm 0.13$ and $\log$([N~II] $\lambda 6583/$H$\alpha) = 0.47 \pm 0.07$, together with $L$([O III] $\lambda 5007$) =  $(1.0\pm 0.1) \times 10^{41}$ ergs s$^{-1}$, 
classify the SDSS spectrum as a type 2 AGN in the LINER region \citep{Kewley2006} in the diagnostic narrow-line diagrams \citep{BPT1981,Veilleux1987, Kauffmann2003} shown in Figure \ref{fig:bpt}.  Note that the archival WISE colors from the all-sky survey in 2010 \citep{Cutri2011} of W1-W2 = 0.48 $\pm 0.04$ mag and W2-W3 = 1.6 $\pm 0.2$  mag, where W1, W2, and W3 are 3.4, 4.6, and 12 $\mu$m, respectively, place the host in the region of Seyfert and star-forming galaxies \citep{Yan2013}.  However, in the WHAN diagram (shown in Figure \ref{fig:bpt}) for emission-line galaxies \citep{Cid2010,Cid2011}, the weak equivalent-width of H$\alpha$ ($W_{\rm H \alpha} = 1.6 \pm 0.3$ \AA) classifies the galaxy as a ``retired galaxy''  powered by hot low-mass evolved (post-asymptotic giant branch) stars and {\it not} an AGN.  

\subsection{Change of State}

The transient iPTF 16bco shows two remarkable changes: a factor of 10 increase in UV flux, and a transformation from a LINER galaxy to a luminous type 1 quasar.  Figure \ref{fig:spec} shows the dramatic change of state between the SDSS spectrum in 2004 and the follow-up Keck spectrum in 2016.    

\subsubsection{Continuum Variability}

According to the empirical correlation between hard X-ray emission and [O III] luminosity for AGNs \citep{Heckman2005, Ueda2015}, from the [O III]$\lambda 5007$ luminosity measured in the pre-event SDSS spectrum in 2004, one would expect $L(2-10 $)keV $\sim 10^{43}$ ergs s$^{-1}$.  This could be even lower if the narrow line emission in iPTF 16bco is powered by stars, and infers a pre-event X-ray luminosity at least an order of magnitude below the X-ray luminosity of $1.5 \times 10^{44}$ ergs s$^{-1}$ measured by Swift XRT during the type 1 state of iPTF 16bco in 2016.   The change in $NUV$ flux between the \textsl{GALEX} AIS measurement in 2004 and the \textsl{Swift} UVOT measurement in 2016 also indicates a brightening by $\Delta m = -2.4 \pm 0.4$ mag.  Allowing for some fraction of the $NUV$ flux in the \textsl{GALEX} measurement to be from star-formation in the host galaxy, this yields a lower-limit of a factor of $\sim 10$ increase in flux.

When adding back in the host galaxy $r$-band magnitude measured by SDSS (see \S \ref{sdss}) the amplitude of optical variability since 2003 is only $0.3$ mag.  However, since the SDSS photometry in 2003 is dominated by host galaxy starlight, the $0.3$ mag of variability is only a {\it lower} limit to the true amplitude of the continuum increase in the optical. 

Thus we conclude that the continuum in iPTF 16bco had an increase in X-ray/UV/optical flux by a factor of $> 10$ between 2004 and 2016.  It is not entirely surprising that such a large increase in the photoionizing continuum was also accompanied by dramatic spectral changes.

\subsubsection{Spectral Variability}\label{sec:lines}

\begin{figure*}
\plottwo{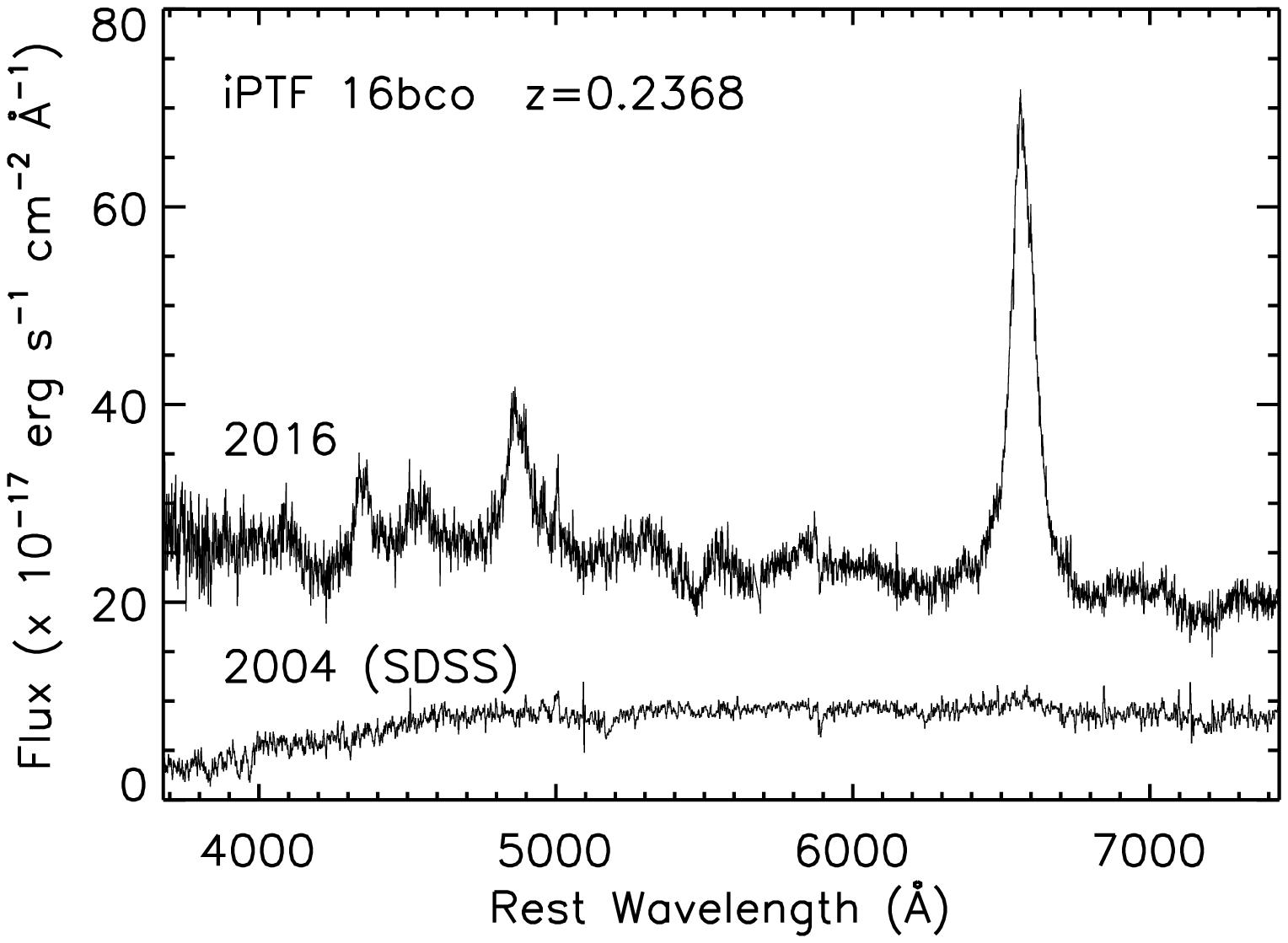}{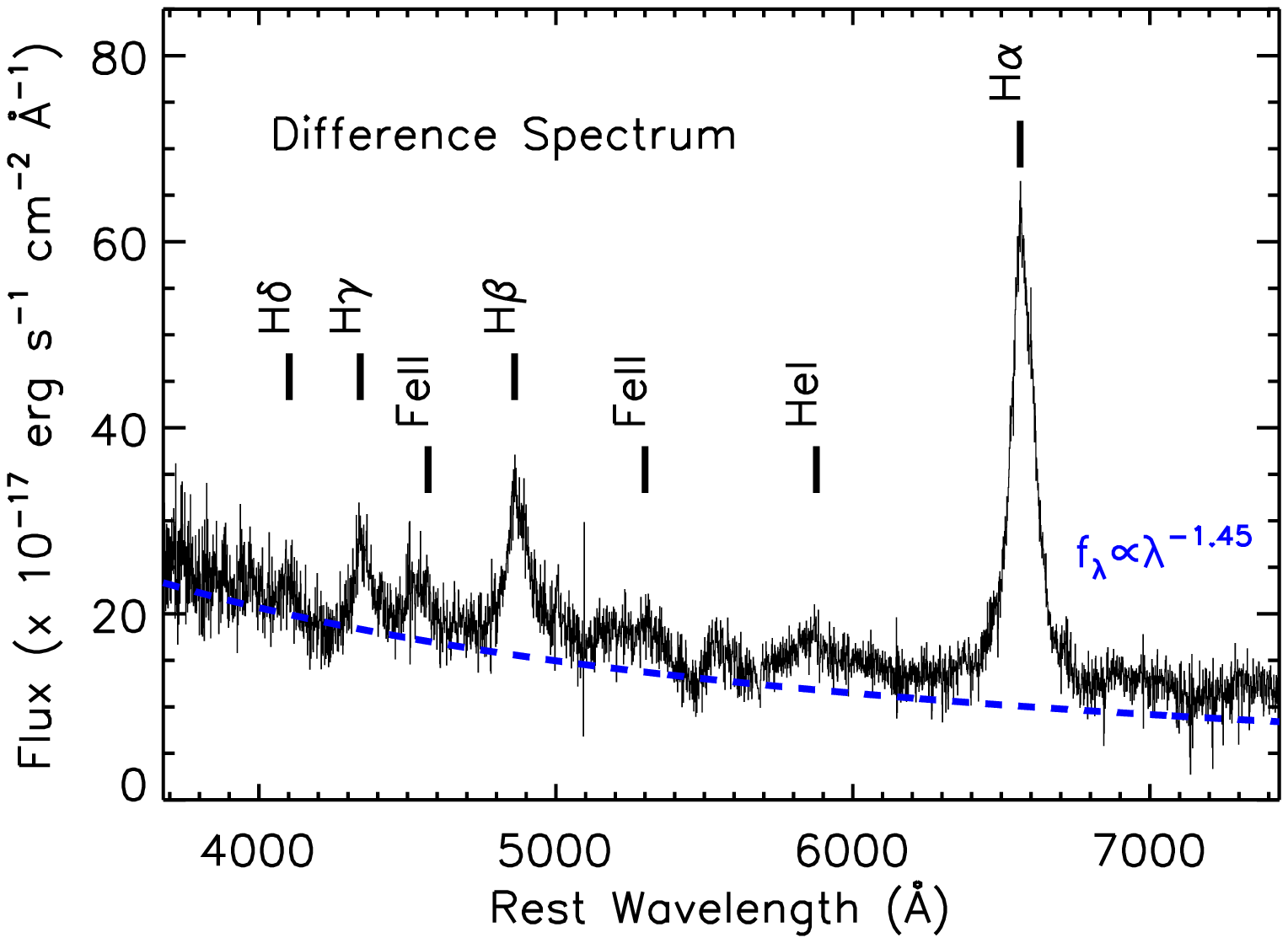}
\caption{{\it Left}: Dramatic change in spectrum between the archival SDSS legacy spectrum obtained on 2004 June 16, and the follow-up spectrum obtained by Keck2+DEIMOS on 2016 June 4.  {\it Right}: Difference spectrum corrected for Galactic extinction.  Tick marks show the broad Balmer lines (H$\alpha$, H$\beta$, and H$\delta$) as well as the broad Fe II complexes, and He I $\lambda 5877$.  Dashed blue line shows a power law fit to the continuum, $f_\lambda \propto \lambda^{\alpha_\lambda}$, with $\alpha_\lambda = -1.45$.
 \label{fig:spec} }
\end{figure*}

The spectra of iPTF 16bco demonstrate a strong blue continuum and broad Balmer-line, Fe II, and He I features characteristic of a type 1 quasar.  Figure \ref{fig:spec} also shows the difference between the Keck spectrum in 2016 and the archival SDSS spectrum from 2004.  After correcting for Galactic extinction of E(B$-$V) = 0.021 mag from the \citet{Schlegel1998} dust extinction map and using the extinction curve of \citet{Cardelli1989}, the continuum in the difference spectrum (after masking the broad emission lines) is reasonably fitted with a single power law of $f_\lambda \propto \lambda^{\alpha_\lambda}$, where $\alpha_\lambda =  -1.45$.  This is shallower than the standard theoretical thin accretion disc spectrum with $f_\nu \propto \nu^{\alpha_\nu}$, where $\alpha_\nu =1/3$, and $\alpha_\lambda = -(\alpha_\nu + 2) = -2.33$, which is well fitted to difference spectra \citep{Wilhite2005, MacLeod2016} and difference spectral energy distributions \citep{Hung2016} of quasars.  However, it is close to the power-law observed in {\it averaged} optical quasar spectra ($\alpha_\lambda = -1.56$) blueward of H$\beta$ \citep{VandenBerk2001, Wilhite2005}.  A power-law index similar to the average quasar spectrum in the difference spectrum is another indication that the low-state spectrum has very little contribution from a non-stellar continuum.

The [O III]$\lambda 5007$ luminosity in 2016 is $(1.6 \pm 0.3) \times 10^{41}$ ergs s$^{-1}$, consistent within the errors of the [O III] luminosity measured in SDSS.   The broad H$\alpha$ flux, in contrast, dramatically appears, with $L$(H$\alpha, {\rm broad}) = (8.22 \pm 0.09) \times 10^{42}$ ergs s$^{-1}$ with a full-width at half-maximum (FWHM) of $4048 \pm 36$ km s$^{-1}$.  This difference in behavior between the broad and narrow lines can be explained by the fact the [O~III] line luminosity traces the average AGN continuum over a much longer timescale that the flaring event detected by iPTF.  While the emissivity decay time (dominated by recombination charge transfer) for [O~III]$\lambda 5007$ is only a few years, light-travel time effects prolong the response time to hundreds of years (the light-crossing time of the narrow-line region) \citep{Eracleous1995}.  Thus the narrow-line region has not yet had the chance to respond to the continuum flux changes happening on the timescale of $\aplt 1$ year.

The monochromatic optical luminosity in the 2016 Keck spectrum after subtracting the starlight using the archival SDSS spectrum, is $\lambda L_\lambda(5100$\AA) = $\lambda f_\lambda 4 \pi d_L^2 (1+z) = 1.6 \times 10^{-16}$ ergs s$^{-1}$ \AA$^{-1} 4 \pi d_L^2 (1+z) = 1.7 \times 10^{44}$ erg s$^{-1}$.  This is in excellent agreement with the nearly linear correlation between broad H$\alpha$ luminosity and optical continuum luminosity in AGN \citep{Greene2005}, which for this broad H$\alpha$ luminosity, one would expect $\lambda L_\lambda $(5100 \AA)$ \sim 1.5 \times 10^{44}$ ergs s$^{-1}$.  

 The FWHM of the broad H$\beta$ line, $4770 \pm 200$ km s$^{-1}$, and the monochromatic luminosity at 5100 \AA~can be used to estimate the central black hole mass to be $2^{+4.0}_{-1.5} \times 10^{8} M_\odot$ \citep{Vestergaard2006}, in good agreement with $M_{\rm BH}$ inferred from the host galaxy stellar velocity dispersion.  We adopt a bolometric correction factor for the monochromatic optical luminosity of 8.1 \citep{Runnoe2012} to get $L_{\rm bol}=1.4 \times 10^{45}$ ergs s$^{-1}$.  We then derive an Eddington ratio of $\lambda_{\rm Edd} = L_{\rm bol}/L_{\rm Edd} = 0.05$ during the type 1 quasar state.  
 
\subsection{Nature of the Variability}

\subsubsection{Variable Obscuration or Microlensing}

The Eddington ratio inferred for iPTF 16fnl from its nearly constant narrow [O~III]$\lambda 5007$ line luminosity 
is in disagreement with the Eddington ratio observed in its type 1 quasar state ($\lambda_{\rm Edd} \sim 0.05$).  Therefore, we favor a change in accretion rate (intrinsic change of state) in the quasar as opposed to variable obscuration (extrinsic change of state) of a non-variable quasar. 

Timescale arguments also disfavor a variable line-of-sight extinction due to the disappearance of an intervening absorber to explain the appearance of the broad H$\alpha$ line in iPTF 16bco.  In the case of an obscuring cloud, the distance between the nucleus and the cloud must be larger than the radius of the broad-line region it is obscuring.  Using the radius-luminosity relation measured for AGN from reverberation mapping studies of broad lines \citep{Bentz2013}, the luminosity of iPTF 16bco would have H$\beta$ broad-line emission with a characteristic radius of $R_{\rm BLR} \sim 45$ days.  Following the argument of \citet{LaMassa2015} (Equation 4), this translates to a crossing time on a circular, Keplerian orbit, $t_{\rm cross} = \Delta \phi / \omega_K$, where $\Delta \phi = $arcsin$ \left (\frac{r_{\rm src}}{r_{\rm orb}} \right )$ is the angular length of the arc, and $\omega_K = \frac{\sqrt{GM}}{2\pi} r^{-3/2}$ is the Keplerian frequency,
which for $r_{\rm orb} > R_{\rm BLR}$ yields $t_{\rm cross}> 15 $yr, which is much longer than the timescale over which the continuum appeared in iPTF 16bco.  

Finally, we note that lensing of a background broad-line quasar by a star in an intervening galaxy \citep{Lawrence2016} is also ruled out, since the redshift of the quasar is the same as the galaxy in its dim state.  

\subsubsection{Tidal Disruption Event}

One mechanism to rapidly increase the mass accretion rate onto a SMBH is to get a new supply of gas from a star that wanders close enough to the SMBH to be torn apart by tidal forces.   In a tidal disruption event (TDE), roughly half of the disrupted stellar debris remains bound to the black hole, falls back onto the SMBH, circularizes through shocks, and is accreted \citep{Rees1988}.   The characteristic timescale of a TDE is the orbital period of the most tightly bound debris, which is given by $\Delta t = 0.35 M_7^{1/2} m_\star^{-1} r_\star^{3/2}$ yr, where $M_7 = M_{\rm BH}/10^7 M_\odot$, $m_\star = M_\star/M_\odot$ and $r_\star = R_\star/R_\odot$ \citep{Lodato2011}.   The  peak mass accretion rate is given by $\dot{M_{\rm acc}} = (1/3) (M_\star/\Delta t)$, and can exceed the Eddington rate ($\dot{M}_{\rm Edd} = 0.2 M_7 (\eta/0.1)^{-1} M_\odot$ yr$^{-1}$, where $\eta$ is the radiative efficiency) for black holes $< 10^{7} M_\odot$.  Following the peak, a TDE has a characteristic $t^{-5/3}$ power-law decay determined by the fallback rate of the stellar debris to pericenter \citep{Rees1988, Phinney1989, Evans1989}.  

However, we point out several issues with interpreting the flaring state of iPTF 16bco with a TDE.  1) For a $M_{\rm BH} \apgt 10^{8} M_\odot$, the tidal disruption radius  of a solar-type star ($R_{\rm T} = R_\star (M_{\rm BH}/M_{\star})^{1/3} = 3.23 \times 10^{13} M_8^{1/3}$ cm) is smaller than the Schwarzschild radius ($R_{\rm S} = 2GM_{\rm BH}/c^2 = 2.95 \times 10^{13} M_8$ cm), and the star crosses the event horizon before being disrupted.  2) The X-ray power-law ($\Gamma = 2.1$) continuum of iPTF-16bco is unlike the extremely soft, thermal X-ray spectra observed in TDEs \citep{Komossa2002, Miller2015}.  3) The light curve of iPTF 16bco shows a complex shape uncharacteristic of a TDE, with month-long plateau, followed by a 2 week rise to another plateau.  4) The broad Balmer lines in iPTF 16bco are narrower and stronger than have been observed in TDEs, and iPTF 16bco does not have strong broad He II $\lambda 4686$ which is characteristic of TDE spectra \citep{Gezari2012, Arcavi2014, Holoien2014, Holoien2016a, Holoien2016b}.  5) Finally, the fact that the broad-line emission and X-ray continuum in iPTF 16bco are consistent with radio-quiet quasars in the ``Eigenvector 1'' parameter space: the FWHM velocity width of the broad H$\beta$ line vs. the ratio of the equivalent widths of the Fe II $\lambda$4570 complex to broad H$\beta$ strength ($R_{\rm FeII}$) \citep{Sulentic2000}, favors a change in $\dot{M_{\rm acc}}$ of a pre-existing accretion disk, instead of a newly formed debris disk from a TDE.  Continued photometric monitoring can determine if the light curve of iPTF 16bco eventually evolves into a power-law decline as expected for a TDE.

\subsection{Accretion Disk Instabilities}

We now investigate the scenario that iPTF 16bco was the result of a change of state in a pre-existing quasar accretion disk.   Interestingly however, iPTF 16bco puts stringent limits on the timescale over which such accretion rate changes must occur.  In the rest-frame, iPTF 16bco demonstrates a dramatic change in continuum flux over a timescale of $\Delta t < 4/(1+z) = 3.23$ yr or in $< 1.1$ yr based on the archival X-ray upper limits.  The timescale by which an accretion disk can change its accretion rate should be determined by the viscous radial inflow timescale, $t_{\rm infl}$.  Furthermore, this timescale is expected to be longer for a quasar ``turning on'' instead of ``turning off'', since it scales as $\lambda_{\rm Edd}^{-2}$ \citep{LaMassa2015}.  The $t_{\rm infl}$ corresponding to the Eddington ratio and black hole mass estimated for iPTF 16bco in its dim state, and assuming a radiation-pressure dominated inner region of a Shakura-Sunyaev disk, is,

\[ t_{\rm infl} = 1300 {\rm yr} \left [\frac{\alpha}{0.1} \right ]^{-1} \left [ \frac{\lambda_{\rm Edd}}{0.005} \right ]^{-2} \left [\frac{\eta}{0.1} \right ]^{2} \left [\frac{r}{10 r_{\rm g}} \right ]^{7/2} \left [\frac{M_8}{2.0} \right ], \]

\noindent a much longer timescale than the observed rapid change in continuum flux in iPTF 16bco.   

One possibility could be an accretion disk eruption as the result of thermal-viscous instabilities in a partial ionization zone, analogous to the outbursts observed in cataclysmic variables and X-ray novae \citep{Siemiginowska1996}.  However, while such instabilities can produce amplitudes of a factor of $10^4$, the expected durations scale with the central mass, and so while cataclysmic variables (CVs) and X-ray novae show large-amplitude outbursts on the timescale of weeks to months, for an accretion disk around a $10^{8} M_{\odot}$ black hole, this corresponds to timescales of $\sim 10^5$ yr.  However, state changes on $\sim$ minute timescales have been observed in some X-ray binaries \citep{Fender2001, Fender2004}, which would imply similar changes in a SMBH on the timescale of only $\sim 10$ yr.

In the accretion-disk instability model, the more narrow the unstable zone, the more rapid the timescale of variability.  The size of the unstable zone depends on the accretion rate:  the lower the accretion rate, the closer in the unstable region is to the inner edge of the disk.  However, the smaller the unstable zone, the smaller the expected amplitude of variability.  For models that demonstrate a factor of $\sim 10$ variability ($\alpha = 0.1, \dot{M}=3 \times 10^{-5} M_\odot$ yr$^{-1}$) in \citet{Siemiginowska1996},  they have a ``turn-on'' timescale of $\sim 10^3$ yr.  This is still 3 orders of magnitude longer than we require for the ``turn-on'' timescale of iPTF 16bco.  

Interestingly, the thermal timescale itself, $t_{\rm th} \sim 1/\alpha \Omega_{\rm K}$, is much shorter, adopting Equation 8 from \citep{Siemiginowska1996}:

\[ t_{\rm th} \sim 2.7 (\alpha/0.1)^{-1} M_8^{-0.5} (r/10^{16} {\rm cm})^{1.5}~ {\rm yr}. \]

\noindent A disk with local thermal fluctuations, potentially driven by the magneto-rotational instability, could be consistent with the rapid timescale of the continuum variability in iPTF 16bco \citep{Dexter2011}.  While an inhomogeneous disk has been demonstrated to fit composite difference spectra \citep{Ruan2014} and color variability of quasars \citep{Schmidt2012}, \citet{Hung2016} find a simple disk model adequate to fit difference-flux UV/optical spectral energy distributions of individual quasars.  Moreover, \citet{Kokubo2015} argue that the tight inter-band correlations observed in SDSS quasar light curves are inconsistent with the inhomogeneous disk model.  However, whether or not these thermal fluctuations can be coherent enough to produce a large-amplitude outburst, is still to be determined.

An instability that arises in a radiation-pressure dominated disk \citep{Lightman1974} has been used to model recurrent flares in accreting Galactic X-ray binaries \citep{Belloni1997} and has recently been applied by \citet{Saxton2015} and \citet{Grupe2015} to explain the large-amplitude soft X-ray flares in Seyfert galaxies NGC 3599 and IC 3599, respectively.  In this scenario, when the internal radiation pressure of the disk become greater than the gas pressure, a heating wave propagates through the disk.   This results in a enhanced local viscosity, scale height, and accretion rate, which rapidly drains the disk.  The instability recurs when the inner disk fills back in.  The rise-time and recurrence timescale can be as short as a year to hundreds of years for a $10^8 M_\odot$ black hole, respectively.

\begin{figure*}
\plottwo{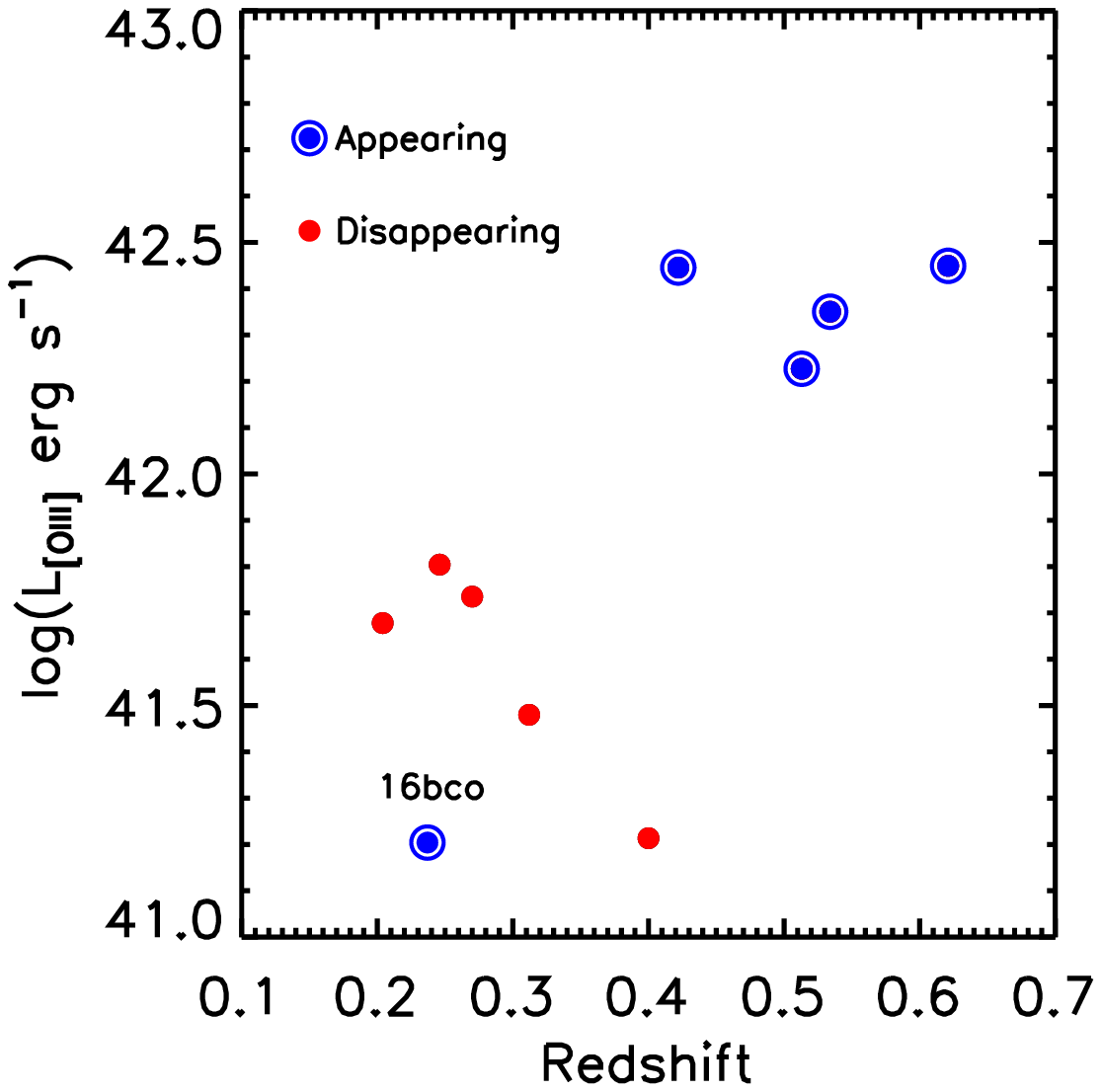}{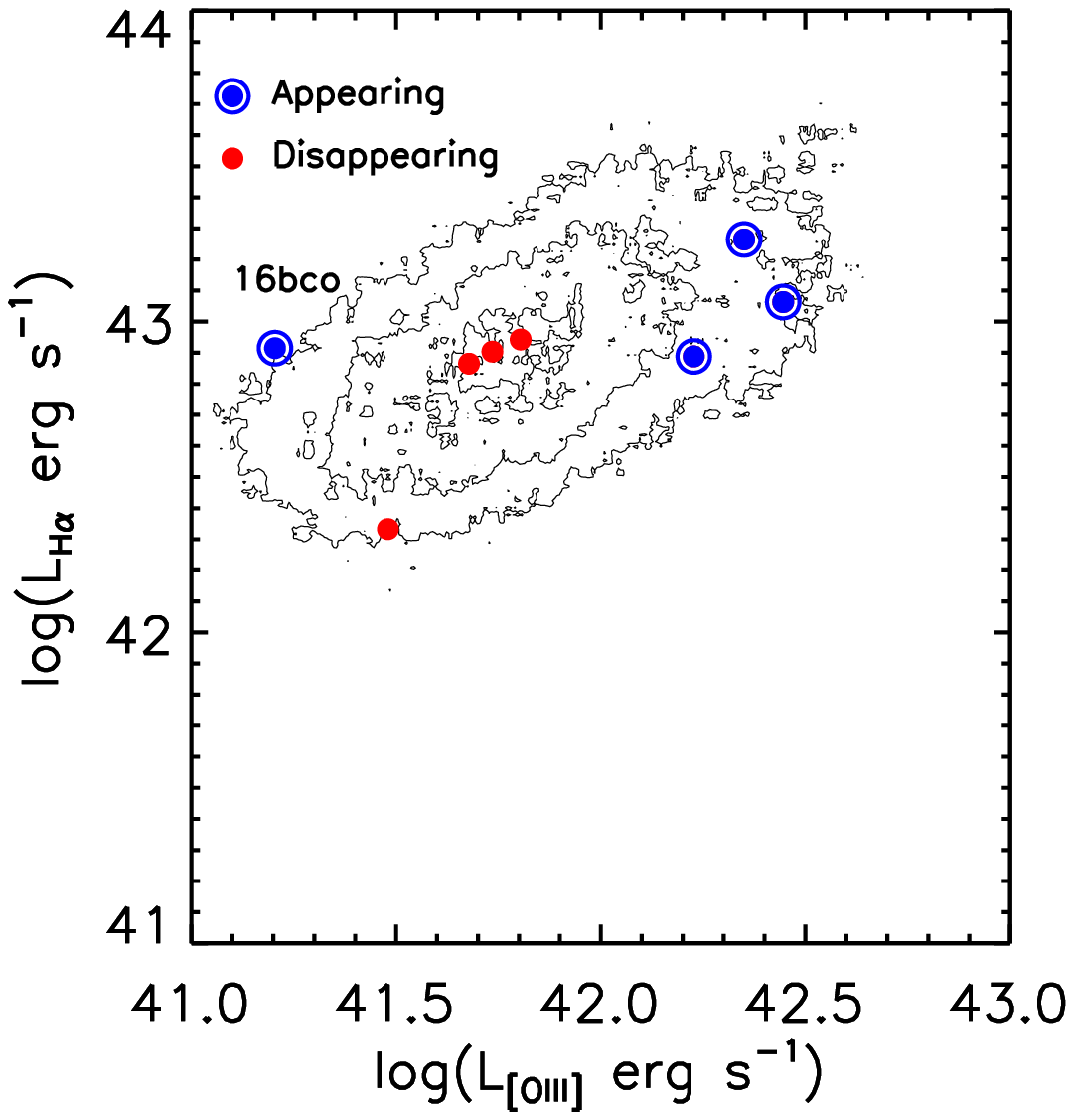}
\caption{Comparison to other changing-look quasars that have disappearing (dots) and appearing (circled dots) broad-line emission.  {\it Left}: [O III]$\lambda 5007$ luminosity during the high-state of the quasar vs. redshift.  {\it Right}: Broad H$\alpha$ luminosity during the high-state of the quasar vs. the [O III]$\lambda 5007$ luminosity during the high-state.  Contours show the distribution of 20\%, 50\%, and 90\% of the quasars' L(H$\alpha$) vs. L([O III]) ratio for the DR7 SDSS quasar sample from \citet{Shen2011}.  iPTF 16bco is an outlier of this distribution, with a high broad H$\alpha$ luminosity relative to its [O III]$\lambda 5007$ luminosity.
 \label{fig:comp} }
\end{figure*}

Another mechanism for driving large amplitude variability in a quasar accretion disk could be related to the presence of a binary SMBH.  Hydrodynamical simulations show that in a circumbinary disk, streams penetrate the disk cavity to feed the primary and secondary black hole at a periodic rate, and that at close to equal mass ratios, the perturbed circumbinary disk has an enhanced accretion rate that can be quite bursty on a timescale of $\sim 5$ times the orbital period of the SMBH binary \citep{Farris2014}.  Such a scenario could in principle be testable, due to the intrinsically periodic nature of the outbursts.  Furthermore, one could look for periodic changes in the broad-line profiles, if they originate in the circumbinary disk.  However, the circumbinary disk outbursts in \citet{Farris2014} have a sawtooth pattern, with a rise time much shorter than the decay time.  Already the light curve of iPTF 16bco is not in good agreement with this model, given its rebrightenning by $\sim 0.5$ mag during its high state (see Figure \ref{fig:lc}).  

The continued intrinsic variability during the high state of iPTF 16bco, as also seen in the study of changing look quasars by \citet{MacLeod2016}, may also provide insight as to the nature of what caused the ``changing look'' of the quasar.  The optical variability amplitudes of these sources in their type 1 quasar states of $\sim 0.5-1.0$ mag are on the high amplitude tail of what is typically observed on these timescales for quasars of a similar luminosity range \citep{MacLeod2012}.  Although we note that in both of our studies the changing look quasars were selected by their optical variability.  However, intrinsic variability is also observed in the changing-look quasar presented in \citet{LaMassa2015}, which was selected based on its spectral changes.  In fact, the rapid rise and power-law decline of its light curve was interpreted as a signature of a TDE \citep{Merloni2015}.  However, the light curves of iPTF 16bco and the other changing-look quasars with archival photometry, have large fluctuations that are inconsistent with the smooth, power-law decline observed in the optical light curves of known tidal disruption events \citep[e.g.,][]{Gezari2008, Velzen2011, Gezari2012, Arcavi2014, Holoien2014, Holoien2016a, Holoien2016b}.
The erratic intra-burst behavior in the two-state (hot and cold) $\alpha$ instability models of \citet{Siemiginowska1996} could be promising.  A potential testable prediction, is that in these models, the accretion disks spend the majority of their time in the low-state.  The variability during the enhanced accretion state in iPTF 16bco could also be a signature of clumpy accretion in an advection-dominated accretion flow (ADAF), for which cold clumps form in the accretion flow due to instabilities in the radiation-dominated regions of the disk \citep{Wang2012}

\subsection{Disk-Jet Connection}

The dramatic change in accretion rate from $\lambda_{\rm Edd} \aplt 0.005$ to  $\lambda_{\rm Edd} \sim 0.05$ inferred for iPTF 16bco could be accompanied by a structural change in the accretion flow if the quasar accretion disk is transitioning from a radiatively inefficient to radiatively efficient mode.  Note that such changes in X-ray binaries are often accompanied by changes in jet activity.
The implied high-state radio-to-optical flux density ratio for iPTF 16bco of $R = \log (S_{\rm 1.4 GHz} /S_{\rm opt}) < 0.8$ and radio luminosity $L_{\rm R} < 3 \times 10^{22}$ W Hz$^{-1}$, is typical of radio-quiet AGN \citep[e.g.,][]{Padovani2011}. While these values are consistent with the LINER classification from the optical spectrum, it is surprising that if the accretion event in iPTF 16bco were triggered by a disk instability, that there is no evidence for a jet or outflow during its high-state in the radio. This is in contrast with X-ray binaries and CVs, which generally show flaring at radio, optical and X-ray wavelengths alongside strong Balmer emission lines. Similarly, the fundamental plane of black hole activity, e.g. \citep{Plotkin2012, Saikia2015} predicts $L_{\rm R}$ significantly greater than $\sim 10^{22}$ W Hz$^{-1}$ when $L_{\rm X} = 1.5 \times 10^{44}$ erg s$^{-1}$. 

\subsection{Comparison to Other Changing-Look Quasars}

iPTF 16bco is one of only a dozen other changing-look quasars (here we define as $M_i < -22$ mag, $L$([O~III])$ > 10^{41}$ ergs s$^{-1}$), roughly half of which have been caught in the act of ``turning on'' by demonstrating the sudden appearance of broad-lines.  Figure \ref{fig:comp} shows the redshift and [O~III] luminosity of all the changing-look quasars in the literature that pass our [O~III] luminosity cut (thus we exclude SDSS J0126-0839 and SDSS J2336+0017 from \citet{Ruan2016}), color-coded by whether they show appearing broad lines, or disappearing broad lines.  We also do not include three changing-look quasars from the \citet{MacLeod2016} sample that do not have good coverage of the broad H$\alpha$ line in its high state (appearing SDSS J214613 at $z=0.62$, disappearing SDSS J022562 at $z=0.63$, and both appearing and disappearing SDSS J022556 at $z=0.50$).  Given that all the other appearing changing-look quasars are from \citet{MacLeod2016}, there appears to be a bias towards finding appearing broad-lines in higher redshift galaxies.  This is likely due to the fact that the BOSS spectra extend to longer wavelengths than the SDSS spectra, and thus finding an appearing broad H$\alpha$ line in the BOSS spectrum is possible at higher-redshift than finding a disappearing broad H$\alpha$ line in the SDSS spectrum. 

Figure \ref{fig:comp} also shows the [O~III]$\lambda 5007$ vs. broad H$\alpha$ luminosity in all the changing-look quasars in their type 1 state, in comparison to the full SDSS quasar sample.  We determine the [O~III]$\lambda 5007$ luminosity from the line flux measured in the SDSS DR7 {\tt SpecLine} table and the DR12 interactive spectrum line measurement table.  We use the broad H$\alpha$ fluxes from the \citet{Shen2011} catalog for broad-line quasars, or from the literature when available.  All luminosities are calculated for our adopted cosmology.  Note that iPTF 16bco is on the edge of the normal quasar distribution, while the other changing-look quasars reported in the literature appear to lie squarely in the distribution of normal quasars in this parameter space.

The enhanced broad H$\alpha$ luminosity observed in iPTF 16bco relative to [O~III] in comparison to normal quasars, as well as the previously discovered changing-look quasars, is likely a signature of its rapid transition to a type 1 state.  As discussed in \S \ref{sec:lines}, given the extended size of the narrow-line region, the [O~III] line will lag in its response to a continuum flare in comparison to the broad emission lines due to light-travel time effects.  Interestingly, in the \citet{MacLeod2016} sample,  the rise-time in the continuum flux of the quasars was serendipitously measured by photometric monitoring to be $\sim 1000$ days.   Given the significantly weaker [O~III] line emission relative to broad H$\alpha$ in iPTF 16bco compared to these objects, this would imply an even shorter ``turn-on'' timescale, in agreement with the inferred ``turn-on'' timescale for the continuum in iPTF 16bco of $\aplt 1$ yr.  

One interesting aspect of changing look quasars is the lack of strong changes in Mg II line emission, despite the dramatic changes in the Balmer lines.  This was explained by \citet{MacLeod2016} as due to the relatively weak responsivity of the Mg II line to continuum flux changes, as has been measured in rest-frame UV reverberation mapping studies \citep{Cackett2015}.  Given that Mg II is a low-ionization line, its weak responsivity can be explained as a consequence of the stratification of the broad-line region \citep{Korista2004}.  Unfortunately, we do not have short-enough wavelength coverage in the archival or follow-up spectra to determine the presence and/or response of the Mg II line during the change of state in iPTF 16bco.  

\section{Summary}

We present the rapid ``turn on'' of a luminous broad-line quasar at $z=0.237$ discovered from its nuclear optical variability in the iPTF survey (iPTF 16bco), and identified as a newly emerged quasar from comparison of follow-up spectroscopy with an archival SDSS spectrum from over a decade earlier which shows LINER narrow-line emission potentially powered by stars.  Pre-event optical, UV, and X-ray imaging indicate that the quasar continuum increased by a factor of $> 10$ on a timescale of $\aplt 1$ yr in the quasar rest-frame.  The broadband properties of iPTF 16bco in its high state are best explained by an intrinsic change of state to a radio-quiet type 1 quasar, than variable obscuration or a TDE.  However, continued monitoring will help further constrain the nature of its rapid brightening.  The dramatic appearance of broad Balmer lines during the high-state of iPTF 16bco, with no significant change in the [O~III]$\lambda$5007 line, is explained as a delayed response of gas in the narrow-line region to the flare in photoionizing continuum due to light-travel time effects.  The enhanced broad H$\alpha$ to narrow [O~III]$\lambda$5007 ratio in iPTF 16bco relative to normal quasars and previously reported changing-look quasars, is further evidence that iPTF 16bco may have demonstrated the most rapid change of state yet observed in a quasar.  iPTF 16bco pushes the limits of accretion disk theory, and may represent a new class of state changes in quasars that will be discovered more routinely in regular monitoring of millions of quasars with the next generation of optical time domain surveys (Zwicky Transient Facility and the Large Synoptic Survey Telescope) together with follow-up spectroscopy triggered by flaring events.\\

\acknowledgements

We thank the anonymous referee for their helpful comments that improved the manuscript.  S.G. is supported in part by NSF CAREER grant 1454816 and NASA Swift Cycle 12 grant NNX16AN85G.  S.G. thanks Mike Koss for help with the X-ray data archives.  These results made use of the Discovery Channel Telescope  at  Lowell  Observatory.   Lowell  is  a  private,  non-profit institution dedicated to astrophysical research and public appreciation of astronomy and operates the DCT in partnership with Boston University, the University of
Maryland, the University of Toledo, Northern Arizona University and Yale University.
The W. M. Keck Observatory is operated as a scientific partnership among the California Institute of Technology, the University of California, and NASA; the Observatory was made possible by the generous financial support of the W. M. Keck Foundation.  This research used resources of the National Energy Research Scientific Computing Center, a DOE Office of Science User Facility supported by the Office of Science of the U.S. Department of Energy under Contract No. DE-AC02-05CH11231.  K.P.M's research is supported by the Oxford Centre for Astrophysical Surveys which is funded through generous support from the Hintze Family Charitable Foundation. The AMI telescope gratefully acknowledges support from the European Research Council under grant ERC-2012- StG-307215 LODESTONE, the UK Science and Technology Facilities Council (STFC) and the University of Cambridge. We thank the AMI staff for scheduling the observations.

\bibliographystyle{fapj}

\bibliography{ms.bib}

\clearpage
\LongTables
\begin{deluxetable}{lcccc}
\tablecaption{iPTF Photometry \label{tab1}}
\tablehead{
\colhead{Telescope+Camera} & \colhead{Filter} & \colhead{MJD} & \colhead{Magnitude} & \colhead{Error}
}
\startdata
P48+CFH12k & $g$ & 57540.336 & 19.66 & 0.07 \\
P48+CFH12k & $g$ & 57540.371 & 19.77 & 0.07 \\
P48+CFH12k & $g$ & 57547.328 & 19.58 & 0.11 \\
P48+CFH12k & $g$ & 57547.359 & 19.55 & 0.07 \\
P48+CFH12k & $g$ & 57551.375 & 19.67 & 0.06 \\
P48+CFH12k & $g$ & 57555.297 & 19.83 & 0.12 \\
P48+CFH12k & $g$ & 57555.324 & 19.69 & 0.12 \\
P48+CFH12k & $r$ & 57540.305 & 19.63 & 0.10 \\
P48+CFH12k & $r$ & 57547.293 & 19.72 & 0.13 \\
P48+CFH12k & $r$ & 57555.270 & 19.63 & 0.12 \\
P60+SEDM & $g$ & 57546.410 & 19.68 & 0.03 \\
P60+SEDM & $g$ & 57547.211 & 19.76 & 0.03 \\
P60+SEDM & $g$ & 57553.465 & 19.92 & 0.05 \\
P60+SEDM & $g$ & 57558.398 & 19.86 & 0.08 \\
P60+SEDM & $g$ & 57564.219 & 19.88 & 0.32 \\
P60+SEDM & $g$ & 57579.262 & 19.43 & 0.02 \\
P60+SEDM & $g$ & 57582.199 & 19.33 & 0.02 \\
P60+SEDM & $g$ & 57584.281 & 19.26 & 0.03 \\
P60+SEDM & $g$ & 57586.215 & 19.24 & 0.03 \\
P60+SEDM & $g$ & 57588.258 & 19.25 & 0.04 \\
P60+SEDM & $g$ & 57590.227 & 19.23 & 0.07 \\
P60+SEDM & $g$ & 57592.266 & 19.20 & 0.02 \\
P60+SEDM & $g$ & 57593.266 & 19.14 & 0.02 \\
P60+SEDM & $g$ & 57594.258 & 19.09 & 0.02 \\
P60+SEDM & $g$ & 57595.258 & 19.13 & 0.02 \\
P60+SEDM & $g$ & 57596.281 & 19.12 & 0.02 \\
P60+SEDM & $g$ & 57599.246 & 19.10 & 0.02 \\
P60+SEDM & $g$ & 57601.211 & 19.06 & 0.02 \\
P60+SEDM & $g$ & 57607.199 & 19.10 & 0.01 \\
P60+SEDM & $g$ & 57611.289 & 19.08 & 0.02 \\
P60+SEDM & $g$ & 57613.254 & 19.14 & 0.03 \\
P60+SEDM & $g$ & 57615.242 & 19.04 & 0.04 \\
P60+SEDM & $g$ & 57618.230 & 19.09 & 0.04 \\
P60+SEDM & $g$ & 57620.223 & 19.03 & 0.04 \\
P60+SEDM & $g$ & 57622.266 & 19.01 & 0.04 \\
P60+SEDM & $g$ & 57624.199 & 19.08 & 0.02 \\
P60+SEDM & $g$ & 57626.188 & 19.11 & 0.02 \\
P60+SEDM & $r$ & 57546.406 & 19.59 & 0.03 \\
P60+SEDM & $r$ & 57547.207 & 19.71 & 0.04 \\
P60+SEDM & $r$ & 57551.395 & 19.88 & 0.03 \\
P60+SEDM & $r$ & 57553.461 & 19.74 & 0.05 \\
P60+SEDM & $r$ & 57558.391 & 19.64 & 0.07 \\
P60+SEDM & $r$ & 57564.215 & 19.72 & 0.02 \\
P60+SEDM & $r$ & 57579.258 & 19.59 & 0.04 \\
P60+SEDM & $r$ & 57582.191 & 19.43 & 0.02 \\
P60+SEDM & $r$ & 57584.277 & 19.32 & 0.04 \\
P60+SEDM & $r$ & 57586.211 & 19.28 & 0.03 \\
P60+SEDM & $r$ & 57588.254 & 19.27 & 0.03 \\
P60+SEDM & $r$ & 57592.262 & 19.26 & 0.02 \\
P60+SEDM & $r$ & 57593.258 & 19.26 & 0.02 \\
P60+SEDM & $r$ & 57594.254 & 19.25 & 0.02 \\
P60+SEDM & $r$ & 57595.254 & 19.21 & 0.02 \\
P60+SEDM & $r$ & 57596.277 & 19.19 & 0.02 \\
P60+SEDM & $r$ & 57599.242 & 19.17 & 0.02 \\
P60+SEDM & $r$ & 57601.207 & 19.16 & 0.02 \\
P60+SEDM & $r$ & 57607.191 & 19.17 & 0.02 \\
P60+SEDM & $r$ & 57609.301 & 19.20 & 0.02 \\
P60+SEDM & $r$ & 57611.285 & 19.19 & 0.02 \\
P60+SEDM & $r$ & 57613.250 & 19.18 & 0.03 \\
P60+SEDM & $r$ & 57615.238 & 19.19 & 0.05 \\
P60+SEDM & $r$ & 57618.223 & 19.08 & 0.06 \\
P60+SEDM & $r$ & 57620.219 & 19.12 & 0.03 \\
P60+SEDM & $r$ & 57622.262 & 19.17 & 0.05 \\
P60+SEDM & $r$ & 57624.191 & 19.15 & 0.02 \\
P60+SEDM & $r$ & 57626.184 & 19.17 & 0.02 \\
P60+SEDM & $i$ & 57546.406 & 19.60 & 0.05 \\
P60+SEDM & $i$ & 57547.211 & 19.73 & 0.05 \\
P60+SEDM & $i$ & 57551.398 & 19.82 & 0.09 \\
P60+SEDM & $i$ & 57553.465 & 19.84 & 0.10 \\
P60+SEDM & $i$ & 57558.395 & 19.69 & 0.06 \\
P60+SEDM & $i$ & 57564.219 & 19.75 & 0.07 \\
P60+SEDM & $i$ & 57579.258 & 19.53 & 0.04 \\
P60+SEDM & $i$ & 57582.195 & 19.39 & 0.03 \\
P60+SEDM & $i$ & 57584.281 & 19.35 & 0.04 \\
P60+SEDM & $i$ & 57586.215 & 19.31 & 0.03 \\
P60+SEDM & $i$ & 57588.254 & 19.30 & 0.04 \\
P60+SEDM & $i$ & 57592.262 & 19.19 & 0.03 \\
P60+SEDM & $i$ & 57593.262 & 19.19 & 0.03 \\
P60+SEDM & $i$ & 57594.254 & 19.20 & 0.03 \\
P60+SEDM & $i$ & 57595.254 & 19.19 & 0.03 \\
P60+SEDM & $i$ & 57596.277 & 19.17 & 0.03 \\
P60+SEDM & $i$ & 57599.242 & 19.19 & 0.02 \\
P60+SEDM & $i$ & 57601.211 & 19.18 & 0.03 \\
P60+SEDM & $i$ & 57607.195 & 19.19 & 0.02 \\
P60+SEDM & $i$ & 57611.285 & 19.22 & 0.04 \\
P60+SEDM & $i$ & 57613.254 & 19.18 & 0.03 \\
P60+SEDM & $i$ & 57615.238 & 19.20 & 0.06 \\
P60+SEDM & $i$ & 57618.227 & 19.12 & 0.04 \\
P60+SEDM & $i$ & 57620.223 & 19.13 & 0.04 \\
P60+SEDM & $i$ & 57622.266 & 19.11 & 0.04 \\
P60+SEDM & $i$ & 57624.195 & 19.17 & 0.02 \\
P60+SEDM & $i$ & 57626.188 & 19.16 & 0.02 \\
P60+GRBCam & $g$ & 57576.348 & 19.50 & 0.03 \\
P60+GRBCam & $r$ & 57576.348 & 19.58 & 0.03 

\enddata
\end{deluxetable}

\end{document}